\documentclass[aps,superscriptaddress, showpacs]{revtex4}
\usepackage{graphicx}
\usepackage{color}
\usepackage{subfig}
\usepackage{wrapfig}

\newcommand{\iz}{I$_\textrm{z}$ }

\begin{document}
\title{Phonons and superconductivity in YbC$_6$ and related compounds}
\author{M. H. Upton}
\affiliation{Advanced Photon Source, Argonne National Laboratory, Chicago, IL USA}
\author{T. R. Forrest}
\affiliation{Department of Physics, University of California at Berkeley, Berkeley, CA USA}
\affiliation{London Centre for Nanotechnology and Department of Physics and Astronomy, University College London, United Kingdom}
\author{A. C. Walters}
\affiliation{European Synchrotron Radiation Facility, Grenoble, France}
\author{C. A. Howard}
\affiliation{London Centre for Nanotechnology and Department of Physics and Astronomy, University College London, United Kingdom}
\author{M. Ellerby}
\affiliation{London Centre for Nanotechnology and Department of Physics and Astronomy, University College London, United Kingdom}
\author{A. H. Said}
\affiliation{Advanced Photon Source, Argonne National Laboratory, Chicago, IL USA}
\author{D. F. McMorrow}
\affiliation{London Centre for Nanotechnology and Department of Physics and Astronomy, University College London, United Kingdom}

\begin{abstract}
The out-of-plane intercalate phonons of superconducting YbC$_6$ have been measured with inelastic x-ray scattering.  Model fits to this data, and previously measured out-of-plane intercalate phonons in graphite intercalation compounds (GICs), reveal surprising trends with the superconducting transition temperature.  These trends suggest that superconducting GICs should be viewed as electron-doped graphite.
\end{abstract}


\maketitle
Interest in superconductivity in graphite intercalation compounds (GICs) has been reignited by the discovery of relatively high transition temperatures in YbC$_6$ (6.5 K) and CaC$_6$ (11.4 K) \cite{Weller, Emery}.
Because neither graphite nor the intercalate exhibit a high transition temperature independently the superconductivity must result from the combination of the graphene and intercalate sheets.
The major question in GIC physics is the nature and strength of the graphene-intercalate interactions, which, despite a great deal of theoretical and experimental work, are still not clearly understood \cite{Weller, Emery, Csanyi, MarkDean, Valla, Pan, Mazin1, Mazin, Walters}.   There are now two competing views of the phenomenon.
Superconductivity may arise mainly from the interaction of the electronic states and in-plane phonons of the electron-doped graphene.  Alternately, it may be a result of coupling between interlayer electronic bands and both in-plane intercalate (I$_\textrm{xy}$) and out-of-plane carbon (C$_\textrm{z}$)  phonons.
The question addressed in this paper is whether the dominant contribution to superconductivity in GICs is phonon coupling with the electron-doped graphite band structure or the intercalate band structure.

Angle resolved photoemission (ARPES) experiments suggest that GICs are best viewed as electron-doped graphite, with the intercalate relegated to the role of electron donor.    Electron doping raises the Fermi level above the Dirac point and thus changes the Fermi surface.  The measured electron-phonon coupling to the in-plane carbon C$_\textrm{xy}$  phonons is sufficient to explain the superconductivity in these compounds \cite{Valla, Pan} although this interpretation of the data is not universally accepted \cite{Calandra_epc, Park}.  Furthermore, Raman measurements show greater electron-phonon interaction with the C$_\textrm{xy}$ phonons than predicted by theory \cite{MarkDean}.
The electron doping of the carbon atoms as a function of T$_\textrm{c}$ has been measured and a largely monotonic dependence found \cite{Pan}.

The alternate view is that the intercalate is more directly involved and that interlayer electrons couple to both I$_\textrm{xy}$ and C$_\textrm{z}$ phonons and that this coupling is responsible for superconductivity \cite{Mauri}.
This model has a number of variations because the origin of the interlayer electronic band is controversial.
The interlayer electrons may come from either the graphite bandstructure \cite{Csanyi},
the intercalate bandstructure \cite{Mauri} or both \cite{Mazin1}.
There is strong experimental evidence that the intercalate atoms must be important to superconductivity in GICs.
In CaC$_6$, the isotope effect for Ca is measured to be $\alpha(\textrm{Ca})=-(d\log T_c/d\log M_{\textrm{Ca}})=0.40$, where $M_{\textrm{Ca}}$ is the mass of the calcium atom \cite{HinksAPS}.
This value is even greater than the substantial calculated value of $\alpha(\textrm{Ca})=0.24$ \cite{Calandra_epc}.    A similarly high value was reported for $\alpha(\textrm{Yb})$ \cite{HinksAPS}.  Furthermore, DFT calculations suggest coupling to the intercalate Fermi surface is responsible for superconductivity in some GICs \cite{Mauri}.

YbC$_6$ is a critical material for understanding GIC superconductivity since it is one of only a few GICs superconduct between 11.4 K and 1 K.
Previous papers have often relied on comparing CaC$_6$ ($\textrm{T}_\textrm{c}=11.4$ K) to GICs which superconduct below 1 K or do not exhibit any superconductivity.
Therefore, understanding superconductivity in YbC$_6$ may provide a missing link in understanding the role of the intercalate in GICs superconductivity.

This paper presents the dispersion of out-of-plane intercalate (I$_\textrm{z}$) phonons in YbC$_6$.
The intercalate \iz phonons  are a suggestive probe of the intercalate-graphene interaction because they reflect the
forces between the intercalate and graphene planes.
Additionally, soft phonons in YbC$_6$ are interesting in their own right because calculations predict a considerable density of Yb states at the Fermi level, suggesting the possibility of superconductivity resulting from electron-phonon coupling between soft Yb phonons and electrons \cite{Mazin_YbC6}.
Model fits to the measured \iz YbC$_6$ phonons may be compared to new fits of a number of previously measured GICs: CaC$_6$ \cite{Upton}, KC$_8$, CsC$_8$ and RbC$_8$ \cite{Magerl}.
Comparing the fitted results from the YbC$_6$ to previously measured compounds shows a surprising trend correlated with the superconducting transition temperature.  These trends suggest that the superconducting transition temperature is a function of charge transfer from the intercalate to the graphene sheets.

\begin{figure}
\subfloat[][]{\label{fig:ResultsUC}\includegraphics[bb=1 1 176 248, height=1.8 in]{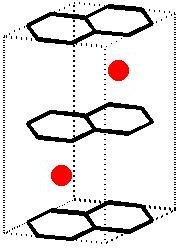}}
\subfloat[][]{\label{fig:ResultsDisp}\includegraphics[bb= 1 1 560 420, width=2.5in]{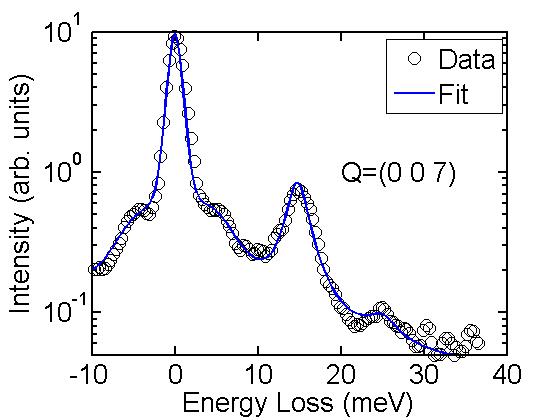}}
\label{fig:Pics}
\caption{(a) The unit cell of YbC$_6$.  Black hexagons represent graphene sheets and red circles represent Yb atoms.  The black dashed
lines represents the unit cell.  The structure can be understood as graphene sheets separated by ordered intercalate layers.
(b) A representative inelastic x-ray spectrum from YbC$_6$.  The peak at zero energy loss is the elastic line.  The features near 15 meV and 25 meV are out-of-plane YbC$_6$ phonons.  The peak near 5 meV is from a phonon branch identified as in-plane YbC$_6$ phonons by comparision with calculations in \cite{Calandra_epc}.
}
\end{figure}

Samples were synthesized using the vapor transport method from natural Madagascan graphite flake, as described in Weller et al.~\cite{Weller}.  The sample dimensions were 3 x 3 x 0.7 mm with a post-intercalation \emph{c}-axis mosaic of $5^\circ$.  The space group of YbC$_6$ is P$6_3$/\emph{mmc}.  The structure is single graphene layers separated by ordered Yb layers, with a=4.32 $\textrm{\AA}$ and c=9.14 $\textrm{\AA}$ \cite{ElMarkini}.
The graphene layers are not staggered, that is, all the carbon atoms in successive layers are superimposed and their stacking is AA.  Successive Yb layers, however, are staggered and have a stacking $\alpha \beta$.  Thus, the entire crystal has a periodicity of A$\alpha$A$\beta$.
A picture of the structure is shown in Fig. \ref{fig:ResultsUC}.

After synthesis, the samples were mounted in a beryllium dome in an argon atmosphere to prevent exposure to oxygen and water, which degrade the samples.  The diffraction pattern of the sample was checked before, after and at several times during the experiment to ensure that their quality did not diminish.  The sample was composed of two regions: a region of YbC$_6$ and an unintercalated graphite region.  The phonons were measured in the region with the largest YbC$_6$ signal which also contained a small amount of unintercalated graphite.  It was possible to exclude graphite signal from this analysis by examining the periodicity of the measured phonons and by comparing the results to the well known graphite phonon dispersion \cite{Mohr}.

Inelastic x-ray  scattering (IXS) experiments were carried out at the high-energy resolution IXS spectrometer (HERIX) at sector 30 at the Advanced Photon Source of Argonne National Laboratory.  The incident synchrotron beam is monochromatized to 1 meV and focused by a Kirkpatrik-Baez mirror to a spot size at the sample of $35 \times 20 \mu\textrm{m}$.  To collect energy loss spectra the incident energy is scanned while the measured energy of the scattered radiation was held constant at 23.724 keV.
Data were collected using Si(12 12 12) analyzer reflections.  The instrument had an overall energy resolution of $\sim1.5$ meV.  Nine analyzer crystals and nine independent detectors allowed data collection at nine momentum transfers simultaneously.  The instrumental momentum resolution was $0.13 \textrm{\AA}^{-1}$.  The spectra were normalized by a beam intensity monitor immediately before the sample.

\begin{figure}
\subfloat[][]{\label{fig:Springmodel}\includegraphics[bb=1 1 626 1098, height=1.8in]{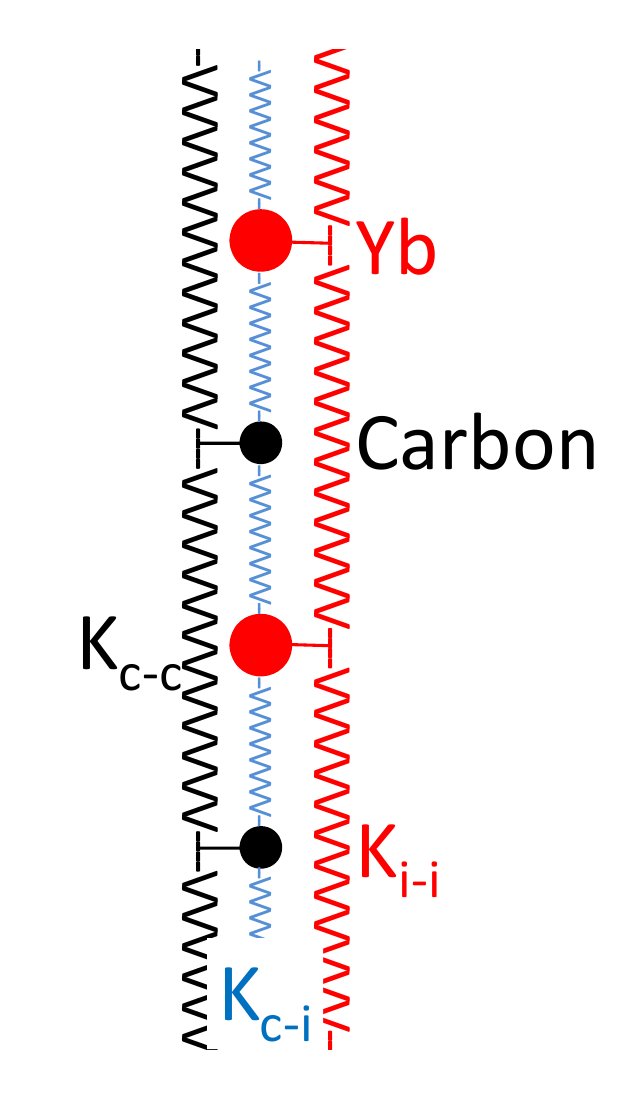}}
\subfloat[][]{\label{fig:SpringFIT}\includegraphics[width=2.5in, bb=1 1 560 420]{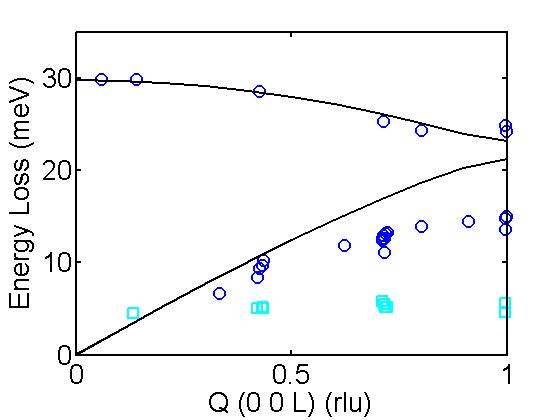}}\\
\subfloat[][]{\label{fig:Resultsmodel}\includegraphics[bb=1 1 849 1099, height=1.8in]{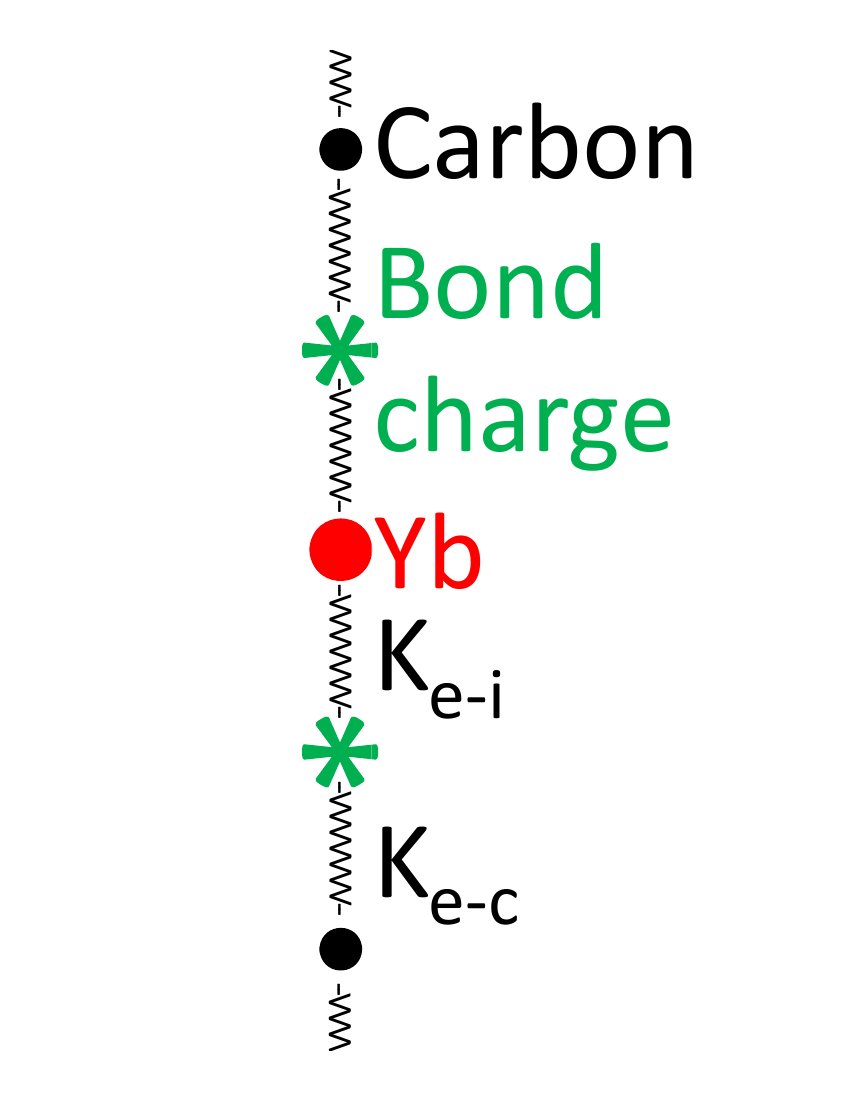}}
\subfloat[][]{\label{fig:ResultsFIT}\includegraphics[width=2.5in, bb=1 1 560 420]{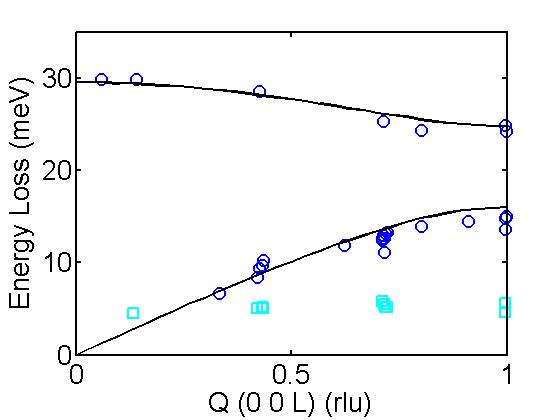}}
\subfloat[][]{\label{fig:Shellmodel}\includegraphics[bb=1 1 1012 1100, height=1.8in]{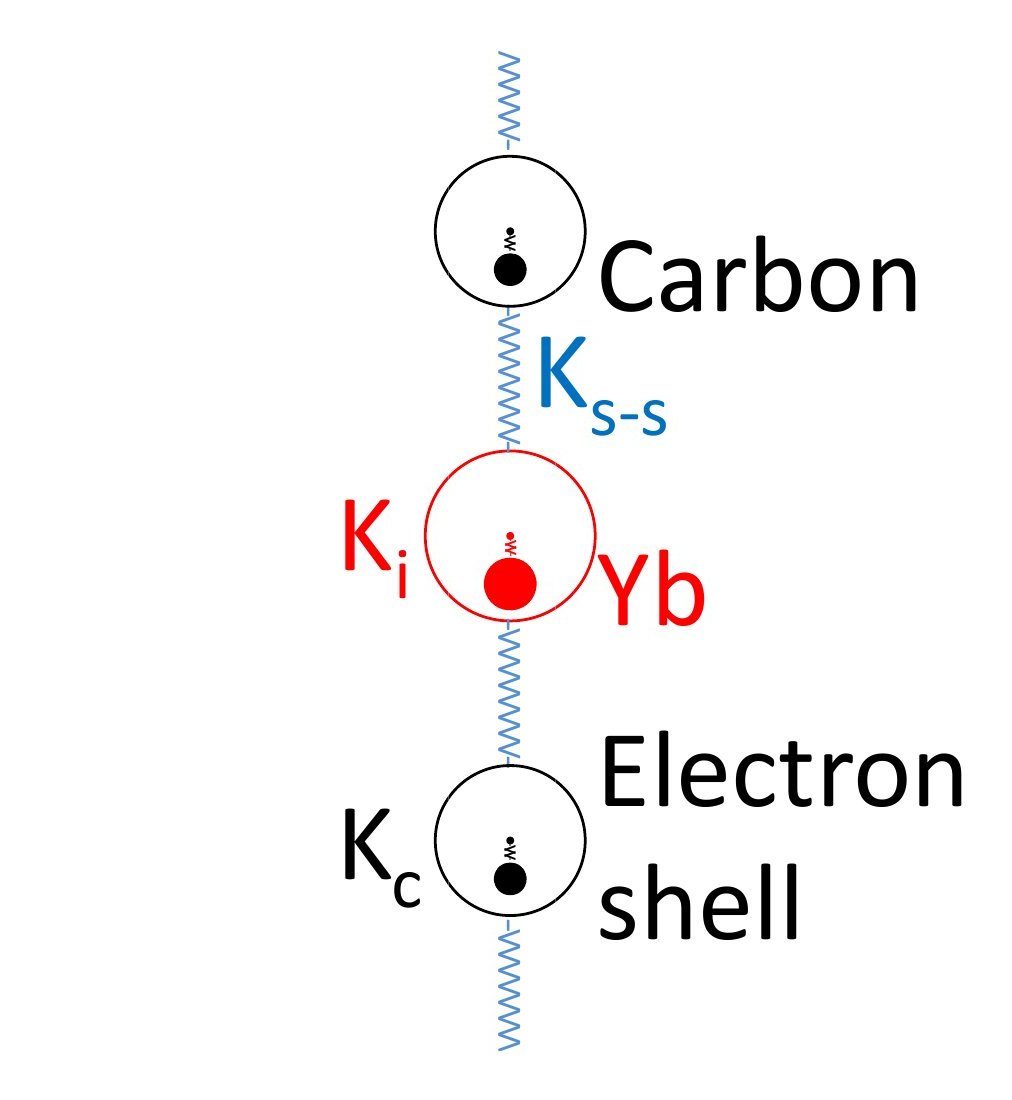}}
\label{fig:Results}
\caption{
(a) A cartoon of the second nearest neighbor model used to fit the phonon dispersion.  Black circles represent the graphene planes; red circles represent the Yb planes. The blue springs connect adjacent
 Yb and graphene planes, the black and red springs connect graphene and Yb sheets to their nearest graphene or Yb neighbors.
(b) [00L] Phonon dispersion in YbC$_6$.  The blue circles represent features assigned to out-of-plane intercalate phonons while the cyan squares represent features assigned to in-plane Yb phonons.
The solid lines are a fit to the out-of-plane phonons with second nearest neighbor springs.
 (c) A cartoon of the bond charge model used to fit the phonon dispersion.
 The green stars represent electrons not strongly bound to either the graphene or Yb atoms.  The spring constant between the electrons and Yb is labeled K$_\textrm{e-i}$ and the spring constant between the electrons and graphene is labeled K$_\textrm{e-c}$.
(d) [00L] Phonon dispersion in YbC$_6$.  The solid lines are a fit to the out-of-plane phonons with the bond charge model.  This fit is indistinguishable from the shell model fit.
(e) A cartoon of the shell model.  The unfilled circles represent the electron shells around the ion cores.  The ion cores are tethered to the center of their
electron shells by springs with spring constants K$_\textrm{i}$ or K$_\textrm{C}$.  The shells are connected to each other with springs with spring constants
K$_\textrm{S-S}$.
}
\end{figure}

A series of spectra were measured at room temperature in reflection geometry, one typical spectrum is shown in Fig. \ref{fig:ResultsDisp}.  The data is shown as black open circles.  A number of spectral features are noticeable.  The large peak at zero energy loss is the elastic line; the three smaller peaks at non-zero energy loss are phonons.
The data is fitted on a log scale to reduce the influence of the elastic line relative to the phonons. A sample fit is shown as a blue line.   The elastic line is fitted as a pseudo-Voigt while each phonon peak is fitted as a Lorentzian.  Corresponding stokes and anti-stokes phonons are constrained to have identical energy losses, widths and the theoretical intensity ratio.  The intensity, width and peak position are fit.
The dispersion of YbC$_6$ phonons is shown in Figs. \ref{fig:SpringFIT} and \ref{fig:ResultsFIT}.  The data are shown as open symbols in a reduced zone scheme with spectra from different Brillouin zones folded back into the first Brillouin zone.  Three bands are measured, two acoustic and one optical.
The spread in measured phonon energies comes from fitting error, inexact sample alignment and integration over the experimental momentum resolution.
By comparison with the calculations of Calandra and Mauri \cite{Calandra_epc} the lower acoustic band, shown as open squares, is attributed to in-plane vibration while the upper acoustic and optical branches, shown as open circles, are attributed to out-of-plane vibration.
It is surprising that the lower acoustic band is observed because the direction of momentum transfer is perpendicular to the predicted direction of vibration and therefore IXS selection rules predict that the phonon would not be observed in this geometry.
Despite this apparent prohibition, however, the lower acoustic mode has been observed in CaC$_6$ \cite{Upton}.  An explanation for this phenomena has been advanced by d'Astuto et al.~\cite{Matteo}.
d'Astuto et al.~have suggested that the lower \iz band does not reach zero at $\Gamma$ in CaC$_6$, but instead interacts with a lower energy mode
and has a finite energy at $\Gamma$ \cite{Matteo}.  Although our data lacks the resolution necessary to observe or eliminate this suggestion, we believe the results of the our model fits are valid.  First, when the optical mode are fit alone, without the acoustic modes, the trends discussed in this paper remain.  Second, the suggestion by d'Astuto, while intriguing, is clearly a higher order effect and involves the details of the interactions in CaC$_6$, not the more general (but less detailed picture) presented in this paper.

In YbC$_6$, no phonon peak widths greater than the instrumental resolution were measured.  Phonon peak broadening beyond the instrumental resolution might have been indicative of electron-phonon coupling, presumably with the interlayer electrons.  The absence of measured broadening does not eliminate the possibility of electron-phonon coupling.

Measurements of the phonon dispersion were repeated at 10 K, just above the superconducting transition temperature. The low temperature dispersion was identical to the dispersion measured at room temperature.  Furthermore, no change in phonon peak width were observed.  Had either effect been observed it would have been suggestive of electron-phonon coupling.

In an attempt to understand the physics of GIC interplanar interactions a number of different models were applied to YbC$_6$ and to previously measured GIC dispersions, as in \cite{Zabel_long}.
Three models are discussed here: a simple spring mode, a bond charge model and a shell model.  The best fit comes from the bond charge model and its implications are discussed.
These models are all one-dimensional and the graphene and intercalate planes are treated as mass densities.  The mass densities are calculated from the atomic masses of C and Yb and knowledge of the structure of the compounds.
In all of these models, the in-plane structure of the GICs is ignored.  While in-plane phonons are, potentially, very important to superconductivity in GICs \cite{Valla, Calandra_epc}, they are ancillary to a model designed to illuminate the interactions between the graphene and intercalate planes.

A simple spring models with nearest neighbor and second nearest neighbor springs did not fit the data well.
In this model, neighboring graphene and intercalate planes are connected by springs and each graphene (Yb) plane is connected to the two nearest graphene (Yb) planes.  In total, three spring constants are fit.  A cartoon of the second nearest neighbor spring model is shown in Fig. \ref{fig:Springmodel} and the fitting results are shown as solid lines in Fig. \ref{fig:SpringFIT}.  In particular the model did not fit the phonon gap at L, the edge of the Brillouin zone, and did not reproduce the acoustic mode dispersion.

A fit to the YbC$_6$ data with the bond charge model is shown as black solid lines in Fig. \ref{fig:ResultsFIT}.
A cartoon of this model is shown in Fig. \ref{fig:Resultsmodel}.
The bond charge model assumes some electrons (the bond charge) bond to both the graphene and intercalate planes.  The bond charge is partially localized away from either plane and both the graphene and intercalate planes are coupled to both planes  by separate springs.
The mass of the bond charge is assumed to be zero (adiabatic approximation).  The fitting parameters are the two spring constants - one between the graphene and bond charge, called K$_\textrm{e-c}$, and one between the intercalate and bond charge, K$_\textrm{e-i}$.
Notice that with just two parameters the model is in excellent agreement with the data.  The bond charge model was previously used to fit GIC phonons \cite{Zabel_long}.

Finally, while the bond charge is shown as physically dividing the space between the graphene and intercalate planes in Fig. \ref{fig:Resultsmodel} the model does not require this exact separation but only requires that the net force from the bond charge attraction be perpendicular to the plane.
It is possible to identify the bond charge with the nearly free electron band, proposed by Mazin and others \cite{Mazin1, Boeri, Csanyi, Mazin_YbC6, Mauri}, however, other possible interpretations exist.

In a simple shell model \cite{Bruesch}, a cartoon of which is shown in Fig. \ref{fig:Shellmodel}, a spherical electronic shell isotropically couples to its rigid ion-core by a spring constant, the shell is coupled to the neighboring electronic shell by an additional spring constant.  The model has three fitting parameters: the spring constant between the intercalate ion core and its electronic shell, K$_\textrm{i}$; the spring constant between the carbon ion core and its electronic shell, K$_\textrm{c}$; and the spring constant between the shells of the intercalate and carbon, K$_\textrm{s-s}$.
  The fit generated by this model is indistinguishable from the fit generated by the bond charge model.  However, because it has three parameters, the quality of shell model fit is lower than quality of the bond charge model fit, a two parameter model.

In general models which allowed electron motion separate from nuclear motion were more successful than models which treated the atoms as points without intermediaries.
This provides supporting, though not conclusive, evidence for the existence of the interlayer band.

\begin{figure}
\subfloat[]{\label{fig:fitEC}\includegraphics[width=3in, bb=1 1 560 420]{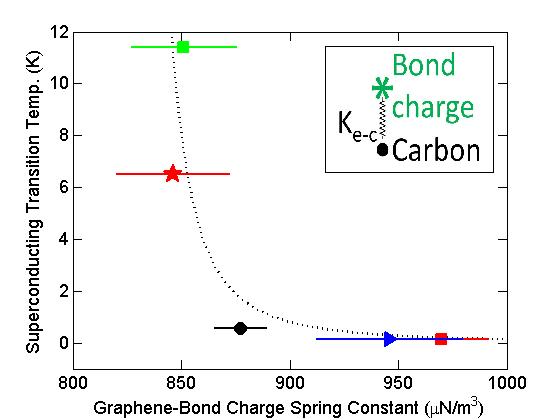}}
\subfloat[]{\label{fig:fitEI}\includegraphics[width=3in, bb=1 1 560 420]{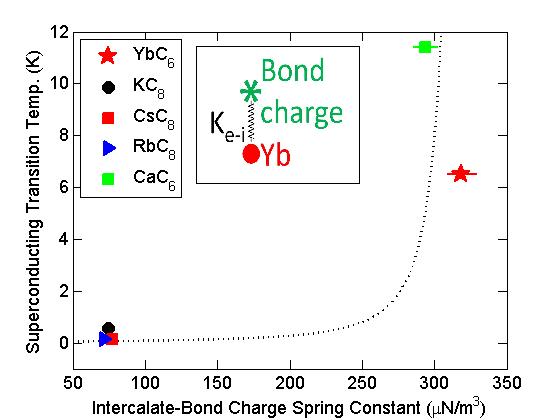}}
\caption{\label{fig:fit}
(a) Fit results from YbC$_6$ and a number of different GICs.  In the first panel, the spring constant connecting the graphene sheets to the bond charge is shown.  The units are $\mu$N/m$^3$ when considering mass per area.  The dotted line is a guide to the eye.
(b) The spring constant connecting the intercalate sheets to the bond charge is shown.  Note that there is over a factor of four difference between the highest and lowest spring constant.  }
\end{figure}

The bond charge model is able to fit the measured dispersions of many  GICs in addition to YbC$_6$, as seen in Fig. \ref{fig:fit} \cite{Upton, Walters, Zabel_long, Magerl, Zabel}, in agreement with the results of Zabel \cite{Zabel_long}.
The fitted spring constants for a series of GICs is shown in Fig. \ref{fig:fit}.
The fitted spring constant between the graphene and electron shell, K$_\textrm{e-c}$, is shown in Fig. \ref{fig:fitEC}.  The overall magnitude of the attraction between the graphene and the bond charge does not change much, but there is a noticeable drop in the superconducting transition temperature value as the spring constant increases.  The fitted spring constant between the intercalate layer and bond charge, K$_\textrm{e-i}$, shown in Fig.~\ref{fig:fitEI}, has a wide range of values.  There is over a factor of four difference between the highest and lowest spring constants.  Higher superconducting transition temperatures are associated with a high intercalate-bond charge spring constant (K$_\textrm{e-i}$).

Both trends in the superconducting transition temperature as a function of spring constant are consistent with the idea of superconductivity coming from electronic doping of the graphene layers by the intercalate.  In the case of greater transfer of electrons from the intercalate to the graphene there will be a stronger attraction (higher spring constant) between the intercalate and the bond charge.  Similarly, if more electrons are transferred to graphene, the attractive interaction between the graphene and bond charge will be reduced.
Previous ARPES measurements also suggested that GICs are best viewed as electron-doped graphene  \cite{Valla,Pan}.
We note, however, that IXS measurements reflect the bulk of the material, while ARPES measurements are sensitive only to the surface.  Therefore, the present measurements provide critical support for this understanding of the system.

The present measurements can be readily compared to previous Raman measurements of Dean et al., which show an interesting, and related, trend linking the out-of-plane carbon phonons and the superconducting transition temperature \cite{MarkDean}.  The zone-center out-of-plane carbon phonons are measured by Raman to be softer in GICs with a high superconducting transition temperature.
These results have been interpreted as showing a correlation between electron-doping of the graphene sheets by the intercalate and the superconducting transition temperature.

%

In conclusion, we have measured the out-of-plane intercalate phonons of YbC$_6$ and fitted them with the bond charge model.
We have applied the model to a series of compounds and, in agreement with previous results \cite{Zabel_long},
found that the bond charge model fits the [00L] intercalate phonons of many first stage GICs very well.  The model fits are consistent with the understanding of superconductivity in GICs arising from phonon coupling to the electron-doped graphene Fermi surface rather than exclusively to the intercalate Fermi surface.  The model can not, however, rule out a contribution to the interlayer band from intercalate atoms.
Additionally, no phonon lifetimes shorter than instrumental resolution were measured in YbC$_6$ at 300 K and no change in the material was observed upon cooling to 10 K.  Although this does not eliminate the possibility of coupling between electrons and intercalate phonons, it does not suggest it either.

Use of the Advanced Photon Source was supported by the U. S. Department of Energy, Office of Science, Office of Basic Energy Sciences, under Contract No. DE-AC02-06CH11357.  The construction of HERIX was partially supported by the NSF under Grant No. DMR-0115852.  Work in London was supported by the EPSRC and a Wolfson Royal Society Award.  We thank Diego Casa for comments on the manuscript.

\bibliography{ArXivYbC6}

\end{document}